\documentclass[showkeys, showpacs, aps,prl, preprint, superscriptaddress]{revtex4}

\usepackage{graphicx}
\usepackage{dcolumn}
\usepackage{bm}
\usepackage{amsmath}
\usepackage{amssymb}
\usepackage{latexsym}
\usepackage{epsfig}
\usepackage{amsbsy}
\usepackage{array}
\usepackage{amssymb}
\usepackage{setspace}
\usepackage{bm}
\usepackage{xcolor}

\def\sint{\ifmmode{- \!\!\!\!\!\! \int}
    \else{\hbox{$- \!\!\!\! \int \ $}}\fi}

\begin{document}

\bibliographystyle{apsrev}
\preprint{Applied Physics Letters}

\title{Spin-current injection and detection in strongly correlated organic conductor}

\author{Z. Qiu\footnote{Author to whom correspondence should be
addressed; electronic mail: qiuzy@imr.tohoku.ac.jp}}
\affiliation{WPI Advanced Institute for Materials Research, Tohoku University, Sendai 980-8577, Japan}
\affiliation{Spin Quantum Rectification Project, ERATO, Japan Science and Technology Agency, Aoba-ku, Sendai 980-8577, Japan}

\author{M. Uruichi}
\affiliation{Institute for Molecular Science, Okazaki 444-8585, Japan.}

\author{D. Hou}
\affiliation{WPI Advanced Institute for Materials Research, Tohoku University, Sendai 980-8577, Japan}
\affiliation{Spin Quantum Rectification Project, ERATO, Japan Science and Technology Agency, Aoba-ku, Sendai 980-8577, Japan}

\author{K. Uchida}
\affiliation{Spin Quantum Rectification Project, ERATO, Japan Science and Technology Agency, Aoba-ku, Sendai 980-8577, Japan}
\affiliation{Institute for Materials Research, Tohoku University, Sendai 980-8577, Japan}
\affiliation{PRESTO, Japan Science and Technology Agency, Saitama 332-0012, Japan}

\author{H. M. Yamamoto}
\affiliation{Research Center of Integrative Molecular Systems (CIMoS), Institute for Molecular Science, 38 Nishigounaka,
Myodaiji, Okazaki 444-8585, Japan.}
\affiliation{RIKEN , 2-1 Hirosawa, Wako 351-0198, Japan. }

\author{E. Saitoh}
\affiliation{WPI Advanced Institute for Materials Research, Tohoku University, Sendai 980-8577, Japan}
\affiliation{Spin Quantum Rectification Project, ERATO, Japan Science and Technology Agency, Aoba-ku, Sendai 980-8577, Japan}
\affiliation{Institute for Materials Research, Tohoku University, Sendai 980-8577, Japan}
\affiliation{Advanced Science Research Center, Japan Atomic Energy Agency, Tokai 319-1195, Japan}


\begin{abstract}
Spin-current injection into an organic semiconductor $\rm{\kappa\text{-}(BEDT\text{-}TTF)_2Cu[N(CN)_2]Br}$ film induced by the spin pumping from an yttrium iron garnet (YIG) film. When magnetization dynamics in the YIG film is excited by ferromagnetic or spin-wave resonance, a voltage signal was found to appear in the $\rm{\kappa\text{-}(BEDT\text{-}TTF)_2Cu[N(CN)_2]Br}$ film. Magnetic-field-angle dependence measurements indicate that the voltage signal is governed by the inverse spin Hall effect in $\rm{\kappa\text{-}(BEDT\text{-}TTF)_2Cu[N(CN)_2]Br}$. We found that the voltage signal in the $\rm{\kappa\text{-}(BEDT\text{-}TTF)_2Cu[N(CN)_2]Br}$/YIG system is critically suppressed around 80 K, around which magnetic and/or glass transitions occur, implying that the efficiency of the spin-current injection is suppressed by fluctuations which critically enhanced near the transitions. 
\end{abstract}



\maketitle


The field of spintronics has attracted great interest in the last decade because of an impact on the next generation magnetic memories and computing devices, where the carrier spins play a key role in transmitting, processing, and storing information \cite{Zutic2004}. Here, a method for direct conversion of a spin current into an electric signal is indispensable. The spin-charge conversion has mainly relied on the spin-orbit interaction, which causes a spin current to induce an electric field $\mathbf{E}_{\rm{ISHE}}$ perpendicular to both the spin polarization $\bm{\sigma}$ and the flow direction of the spin current $\mathbf{J}_{\rm{s}}$: $\mathbf{E}_{\rm{ISHE}}\Vert\mathbf{J}_{\rm{s}}\times\bm{\sigma}$. This phenomenon is known as the inverse spin Hall effect (ISHE) \cite{Saitoh2006,Azevedo2005}. To investigate spin-current physics and realize large spin-charge conversion, the ISHE has been measured in various materials ranging from metals and semiconductors to an organic conjugated polymer \cite{Qiu2012,Ando2010a,Qiu2013,Ando2011a,Qiu2013a,Kajiwara2010, Ando2013, Watanabe2014}. A recent study has suggested that conjugated polymers can work as a spin-charge converter \cite{Ando2013, Watanabe2014}, and further investigation of the ISHE in different organic materials is now necessary. 

\begin{figure}
\centering
\begin{minipage}[b]{0.5\textwidth}
\centering
\includegraphics[width=3in]{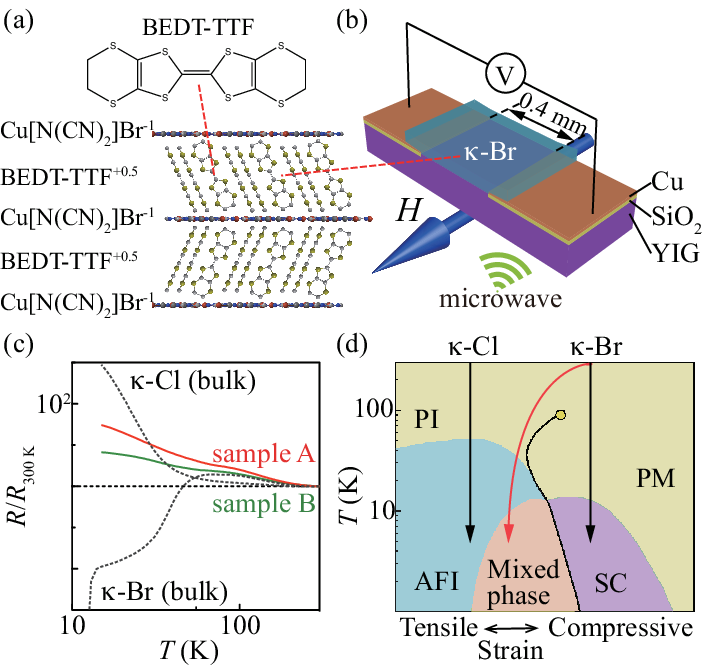}
\end{minipage}
\caption{(a) Structural formula of the BEDT-TTF molecule (upper panel) and schematic cross-section of the (BEDT-TTF)$_2$Cu[N(CN)$_2$]Br ($\kappa$-Br) crystal, where cationic BEDT-TTF  and anionic Cu[N(CN)$_2$]Br layers alternate each other (lower panel). (b) Schematic illustration of the sample structure and experimental setup. $H$ denotes the static external maganetic field applied along the film plane. (c) Temperature dependence of $R/R_{\rm 300 K}$ of the two $\kappa$-Br/YIG samples A and B, a bulk $\kappa$-Br crystal, and a bulk $\kappa$-Cl crystal. Here, $R$ ($R_{\rm 300 K}$) denotes the resistance between the ends of the $\kappa$-Br film at each temperature (at 300 K). (d) Conceptual phase diagram of $\kappa$-X systems. PI, PM, AFI, and SC denote paramagnetic insulator, paramagnetic metal, antiferromagnetic insulator, and superconductor, respectively. The red arrow indicates the trajectory that the $\kappa$-Br crystal on the YIG substrate experiences upon cooling.
\label{figure1}}
\end{figure}

In the present study, we report the observation of the ISHE in an organic molecular semiconductor $\rm{\kappa\text{-}(BEDT\text{-}TTF)_2Cu[N(CN)_2]Br}$ (called $\kappa$-Br). The $\kappa$-Br consists of alternating layers of conducting sheets (composed of BEDT-TTF dimers) and insulating sheets (composed of $\rm{Cu[N(CN)_2]Br}$ anions), which is recognized as an ideal system with anisotropic strongly correlated electrons (Fig. \ref{figure1}(a)). The ground state of bulk $\kappa$-Br is known to be superconducting with the transition temperature $T_c \approx 12$ K\cite{Kini1990}, which becomes antiferromagnetic and insulating by replacing $\rm{Cu[N(CN)_2]Br}$ with $\rm{Cu[N(CN)_2]Cl}$ (Fig. \ref{figure1}(d))\cite{Williams1990}.

To inject a spin current into the $\kappa$-Br, we employed a spin-pumping method by using $\kappa$-Br/yttrium iron garnet (YIG) bilayer devices. In the $\kappa$-Br/YIG bilayer devices, magnetization precession motions driven by ferromagnetic resonance (FMR) and/or spin-wave resonance (SWR) in the YIG layer inject a spin current across the interface into the conducting $\kappa$-Br layer in the direction perpendicular to the interface\cite{Sandweg2010}. This injected spin current is converted into an electric field along the $\kappa$-Br film plane if $\kappa$-Br exhibits ISHE. 

The preparation process for the $\kappa$-Br/YIG bilayer devices is as follows. A single-crystalline YIG film with the thickness of $5\ \rm{\mu m}$ was put on a gadolinium gallium garnet wafer by a liquid phase epitaxy method. The YIG film on the substrate was cut into a rectangular shape with the size of $3 \times 1 ~\textrm{mm}^2$. Two separated Cu electrodes with the thickness of 50 nm were then deposited near the ends of the YIG film. The distance between the two electrodes was 0.4 mm. Here, to completely avoid the spin injection into the Cu electrodes, 20-nm-thick SiO$_2$ films were inserted between the electrodes and the YIG film, as shown in Fig. \ref{figure1}(b). Finally, a laminated $\kappa$-Br single crystal, grown by an electrochemical method \cite{Yamamoto2013}, was placed on the top of the YIG film between the two Cu electrodes (Fig. \ref{figure1}(b)). We prepared two $\kappa$-Br/YIG samples A and B to check reproducibility. The thicknesses of the $\kappa$-Br films for the samples A and B are around 100 nm but a little different from each other, resulting in the difference of the resistance of the $\kappa$-Br film (Fig. \ref{figure1}(b)). To observe the ISHE in $\kappa$-Br induced by the spin pumping, we measured the $H$ dependence of the microwave absorption and DC electric voltage between the electrodes at various temperatures with applying a static magnetic field $H$ and a microwave magnetic field with the frequency of 5 GHz to the device. 


Figure \ref{figure1}(c) shows the temperature dependence of resistance of the $\kappa$-Br films on YIG for the samples A and B. The $\kappa$-Br films in the present system exhibit no superconducting transition \cite{Kini1990, Kanoda2006}, but do insulator-like behavior similar to a bulk $\kappa$-Cl \cite{Williams1990, Kanoda2006}. This result can be ascribed to tensile strain induced by the substrate due to the different thermal expansion coefficients of $\kappa$-Br and YIG. The similar phenomenon was reported in $\kappa$-Br on a $\rm{SrTiO_3}$ substrate\cite{Yamamoto2013}, of which the thermal expansion coefficient is close to that of YIG ($\sim~10~\textrm{ppm/K}$ at room temperature\cite{Ling1993, Boudiar2004a}). Thus, the ground state of the $\kappa$-Br film on YIG is expected to be slightly on the insulator side of the Mott transition. The red arrow in Fig. \ref{figure1}(d) schematically indicates a state trajectory of our $\kappa$-Br films with decreasing the temperature \cite{Miyagawa2004,Kanoda2006,Faltermeier2007,Kagawa2009,Kagawa2005,Limelette2003,Fournier2003,Kuwata2011,Yamamoto2013}.

\begin{figure}
\centering
\begin{minipage}[b]{0.5\textwidth}
\centering
\includegraphics[width=3in]{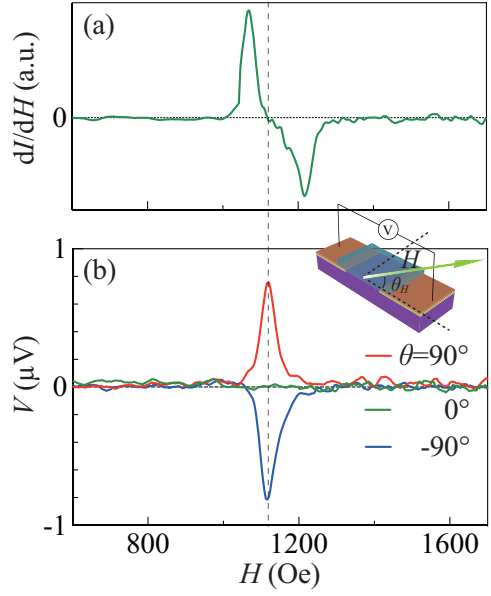}
\end{minipage}
\caption{(a) The FMR/SWR spectrum $dI/dH$ of the $\kappa$-Br/YIG sample A at 300 K. Here, $I$ and denotes the microwave absorption intensity. The dashed line shows the magnetic field $H_{\rm{FMR}}$ at which the FMR is excited. (b) The electric voltage $V$ between the ends of the $\kappa$-Br film as a function of $H$. 
\label{figure2}}
\end{figure}

Figure \ref{figure2}(a) shows the FMR/SWR spectrum $dI/dH$ for the $\kappa$-Br/YIG sample A at 300 K. Here, $I$ denotes the microwave absorption intensity. The spectrum shows that the magnetization in the YIG film resonates with the applied microwave around the FMR field $H_{\rm{FMR}}\approx 1110$ Oe. As shown in Fig. \ref{figure2}(b), under the FMR/SWR condition, electric voltage with peak structure was observed between the ends of the $\kappa$-Br film at $\theta=\pm 90^\circ$, where $\theta$ denotes the angle between the $H$ direction and the direction across the electrodes (Fig. \ref{figure2}(b)). The voltage signal disappears when $\theta=0^\circ$. This $\theta$ dependence of the peak voltage is consistent with the characteristic of the ISHE induced by the spin pumping. Because the SiO$_2$ film between the Cu electrode and the YIG film blocks the spin-current injection across the Cu/YIG interfaces, the observed voltage signal is irrelevant to the ISHE in the Cu electrodes. The magnitude of the electric voltage is one or two orders of magnitude smaller than that in conventional Pt/YIG devices \cite{Iguchi2012,Iguchil2012,Rezende2013,Wang2011}, but is close to that observed in polymer/YIG devices \cite{Ando2013}. 

\begin{figure}
\centering
\begin{minipage}[b]{0.5\textwidth}
\centering
\includegraphics[width=3in]{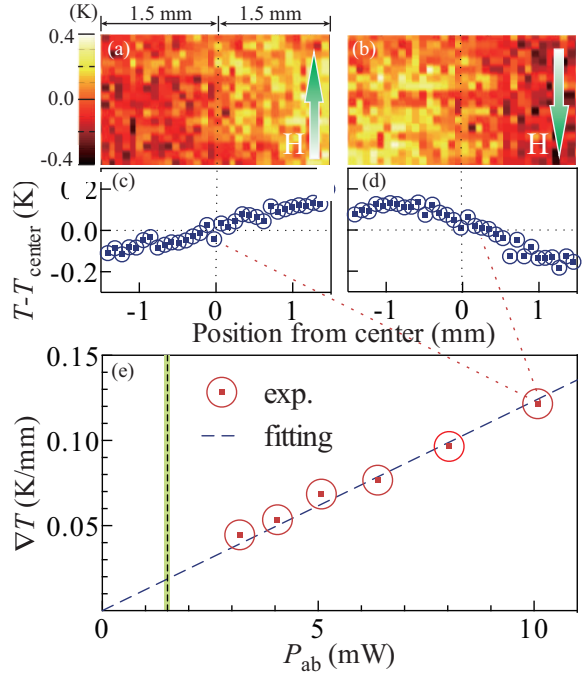}
\end{minipage}
\caption{(a),(b) Temperature distributions of the YIG surface near the FMR fields (5 GHz) for the opposite orientations of $H$, measured with an infrared camera. (c),(d) Temperature profiles of the YIG surface. (e) The microwave-power absorption $P_{\rm ab}$ dependence of the temperature gradient $\nabla T$ of the YIG surface. The voltage measurements were carried out with a low $P_{\rm ab}$ value (marked with a green line). 
\label{figure3}}
\end{figure}

To establish the ISHE in the $\kappa$-Br/YIG sample exclusively, it is important to separate the spin-pumping-induced signal from thermoelectric voltage induced by temperature gradients generated by nonreciprocal surface-spin-wave excitation \cite{An2013}, since thermoelectric voltage in conductors whose carrier density is low, such as $\kappa$-Br, may not be negligibly small. In order to estimate temperature gradient under the FMR/SWR condition, we excited surface spin waves in a 3-mm-length YIG sample by using a microwave of which the power is much higher than that used in the present voltage measurements, and measured temperature images of the YIG surface with an infrared camera (Figs. \ref{figure3}(a) and (b)). We found that a temperature gradient is created in the direction perpendicular to the $H$ direction around the FMR field and its direction is reversed by reversing $H$, consistent with the behavior of the spin-wave heat conveyer effect (Fig. \ref{figure3}(c) and (d)) \cite{An2013}. Figure \ref{figure3}(e) shows that the magnitude of the temperature gradient is proportional to the absorbed microwave power. This temperature gradient might induce an electric voltage due to the Seebeck effect in $\kappa$-Br with the similar symmetry as the ISHE voltage. However, the thermoelectric voltage is expected to be much smaller than the signal shown in Fig. \ref{figure2}(b); since all the measurements in this work were carried out with a low microwave-absorption power (marked with a green line in Fig. \ref{figure3}(e)), the magnitude of the temperature gradient in the $\kappa$-Br/YIG film is less than 0.015 K/mm. Even when we use the Seebeck coefficient of $\kappa$-Br at the maximum value reported in previous literatures \cite{Yu1991,LOGVENOV1995,Yefanov2003}, the electric voltage due to the Seebeck effect in the $\kappa$-Br film is estimated to be less than $0.01 \mu$V at 300 K, where the effective length of $\kappa$-Br is 0.4 mm. This is at least one order of magnitude less than the signals observed in our $\kappa$-Br/YIG sample. Therefore, we can conclude that the observed electric voltage with the peak structure is governed by ISHE.

\begin{figure}
\centering
\begin{minipage}[b]{0.5\textwidth}
\centering
\includegraphics[width=3in]{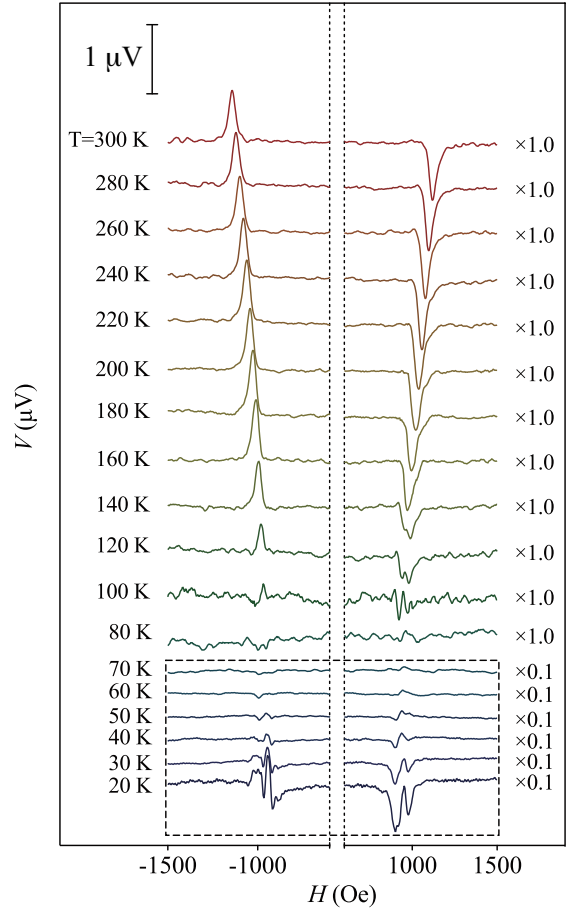}
\end{minipage}
\caption{$H$ dependence of $V$ in the $\kappa$-Br/YIG sample A for various values of the temperature $T$. The scales of the longitudinal axis for the data at $T \leq 70~\textrm{K}$ are shrinked by a factor of 0.1. 
\label{figure4}}
\end{figure}

Figure \ref{figure4} shows the electric voltage spectra in the $\kappa$-Br/YIG sample A for various values of temperature, $T$. Clear voltage signals were observed to appear at the FMR fields when $T > 80~\textrm{K}$. We found that the sign of the voltage signals is also reversed when $H$ is reversed, which is consistent with the ISHE as discussed above. Surprisingly, however, the peak voltage signals at the FMR fields decrease steeply with decreasing $T$ and merge into noise around 80 K. This anomalous suppression of the voltage signals cannot be explained by the resistance $R$ change of the $\kappa$-Br film because no remarkable $R$ change was observed in the same temperatures (Fig. \ref{figure1}(c)). At temperatures lower than 60 K, large voltage signals appear around the FMR fields as shown in Fig. \ref{figure4}, but its origin is not confirmed because of the big noise and poor reproducibility. Therefore, hereafter we focus on the temperature dependence of the voltage signals above 80 K. 

\begin{figure}
\centering
\begin{minipage}[b]{0.5\textwidth}
\centering
\includegraphics[width=3in]{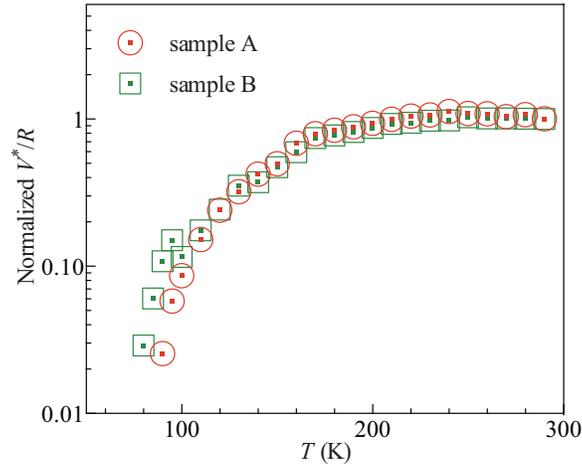}
\end{minipage}
\caption{$T$ dependence of $V^*/R$ for the $\kappa$-Br/YIG samples A and B. Here, $V^*={ \left( {V}_{FMR(-H)}-{V}_{FMR(+H)} \right) }/{ 2 }$ with ${V}_{FMR(\pm H)}$ being the electric voltage at $H_{\rm{FMR}}$. \label{figure5}}
\end{figure}

In Fig. \ref{figure5}, we plot the $T$ dependence of $V^*/R$ for the $\kappa$-Br/YIG samples A and B, where $V^*={ \left( {V}_{{\rm FMR}(-H)}-{V}_{{\rm FMR}(+H)} \right) }/{ 2 }$ with ${V}_{{\rm FMR}(\pm H)}$ being the electric voltage at the FMR fields, to take into account the resistance difference of the $\kappa$-Br films. $V^*/R$ for both the samples exhibit almost same $T$ dependence, indicating that the observed voltage suppression is an intrinsic phenomenon in the $\kappa$-Br/YIG samples.

Here we discuss a possible origin of the observed temperature dependence of the voltage in the $\kappa$-Br/YIG systems. ISHE voltage is determined by two factors. One is spin-to-charge conversion efficiency, i.e. the spin-Hall angle, in the $\kappa$-Br film. The mechanism of ISHE consists of intrinsic contribution due to spin-orbit coupling in the band structure and extrinsic contribution due to the impurity scattering \cite{Vignale2009}. In organic systems such as $\kappa$-Br, the extrinsic contribution seem to govern the ISHE since intrinsic contribution is expected to be weak because of their carbon-based light-element composition. Judging from the predicted rather weak temperature dependence of impurity scattering, the temperature dependence of the spin-Hall angle can not be the origin of the sharp suppression of the voltage signal in the $\kappa$-Br/YIG systems (Fig. \ref{figure5}). The other factor is the spin-current injection efficiency across the $\kappa$-Br/YIG interface, which can be affected by spin susceptibility \cite{Ohnuma2014} in $\kappa$-Br. Importantly, the temperature dependence of the spin susceptibility for the $\kappa$-X family was shown to exhibit a minimum at temperatures similar to those at which the anomalous suppression of the spin-pumping-induced ISHE voltage was observed \cite{Kanoda2006}, suggesting an importance of the temperature dependence of the spin-current injection efficiency in the $\kappa$-Br/YIG systems. We also mention that the temperature at which the ISHE suppression was observed coincides with a glass transition temperature of $\kappa$-Br films  \cite{Strack2005,Muller2002}. However, at the present stage, there is no framework to discuss the relation between the spin-current injection efficiency and such lattice fluctuations. To obtain the full understanding of the temperature dependence of the spin-pumping-induced ISHE voltage in the $\kappa$-Br/YIG systems, more detailed experimental and theoretical studies are necessary.

In summary, we have investigated the spin pumping into organic semiconductor $\rm{\kappa\text{-}(BEDT\text{-}TTF)_2Cu[N(CN)_2]Br}$ ($\kappa$-Br) films from adjacent yttrium iron garnet (YIG) films. The experimental results show that an electric voltage is generated in the $\kappa$-Br film when ferromagnetic or spin-wave resonance is excited in the YIG film. Since this voltage signal was confirmed to be irrelevant to extrinsic temperature gradients generated by spin-wave excitation and the resultant thermoelectric effects, we attribute it to the inverse spin Hall effect in the $\kappa$-Br film. The temperature-dependent measurements reveal that the voltage signal in the $\kappa$-Br/YIG systems is critically suppressed around 80 K, implying that this suppression relates with the spin and/or lattice fluctuations in $\kappa$-Br.


This work was supported by PRESTO ``Phase Interfaces for Highly Efficient Energy Utilization'', Strategic International Cooperative Program ASPIMATT from JST, Japan, Grant-in-Aid for Young Scientists (A) (25707029), Grant-in-Aid for Young Scientists (B) (26790038), Grant-in-Aid for Challenging Exploratory Research (26600067), Grant-in-Aid for Scientific Research (A) (24244051), Grant-in-Aid for Scientific Research on Innovative Areas ``Nano Spin Conversion Science'' (26103005) from MEXT, Japan, NEC Corporation, and NSFC. \par








\end{document}